# Origin of hydrogen isotopic variations in chondritic water and organics

Laurette Piani[1,*], Yves Marrocchi[1], Lionel G. Vacher[2], Hisayoshi Yurimoto[3], Martin Bizzarro[4]

[1]Université de Lorraine, CNRS, CRPG, UMR 7358, Vandoeuvre les Nancy, France

[2]Laboratory for Space Sciences and the Department of Physics, Washington University in St. Louis, St. Louis, MO 63130, USA

[3]Department of Natural History Sciences, Faculty of Science, Hokkaido University, Sapporo, Japan

[4]StarPlan - Centre for Star and Planet Formation, GLOBE Institute, University of Copenhagen, Øster Voldgade 5-7, Copenhagen, Denmark, DK-1350

[*]Corresponding author: laurette.piani@univ-lorraine.fr

**Abstract:**

Chondrites are rocky fragments of asteroids that formed at different times and heliocentric distances in the early solar system. Most chondrite groups contain water-bearing minerals, attesting that both water-ice and dust were accreted on their parent asteroids. Nonetheless, the hydrogen isotopic composition (D/H) of water in the different chondrite groups remains poorly constrained, due to the intimate mixture of hydrated minerals and organic compounds, the other main H-bearing phase in chondrites. Building on our recent works using *in situ* secondary ion mass spectrometry analyses, we determined the H isotopic composition of water in a large set of chondritic samples (CI, CM, CO, CR, CY, and C-ungrouped carbonaceous chondrites) and report that water in each group shows a distinct and unique D/H signature. Based on a comparison with literature data on bulk chondrites and their water and organics, our data do not support a preponderant role of parent-body processes in controlling the D/H variations among chondrites. Instead, we propose that the water and organic D/H signatures were mostly shaped by interactions between the protoplanetary disk and the molecular cloud that episodically fed the disk over several million years. Because the preservation of D-rich interstellar water and/or organics in chondritic materials is only possible below their respective sublimation temperatures (160 and 350–450 K), the H isotopic signatures of chondritic materials depend on both the timing and location at which their parent body formed.

**Keywords:**

chondrite, water, organic matter, hydrogen isotopes, disk, molecular cloud

1. Introduction

Molecular clouds correspond to cold regions (i.e., 10–30 K) of accumulated interstellar gas (mainly $H_2$) and dust in the narrow midplanes of galactic disks. Their isothermal gravitational collapse and fragmentation drive the formation of clusters of tens to several hundreds of protostars; the Sun itself was likely born near a few hundred stars (Gounelle and Meynet, 2012). The conservation of angular momentum induces the rapid formation of protoplanetary disks around young stellar objects, through which materials from molecular clouds are channeled and reprocessed (Pignatale et al., 2018). Protoplanetary disks are expected to rapidly assemble from the infalling molecular clouds, reaching their maximum masses within 300 kyr (Williams and Cieza, 2011; Yang et al., 2013). In contrast, the lifetimes of star-forming molecular clouds are estimated to be ~20–30 Myr (Murray, 2011), and recent hydrodynamic simulations suggest that they can fuel the formation of protoplanetary disks episodically throughout their evolution (Kuznetsova et al., 2020).

Chondrites are rocks leftover from the evolution of the solar protoplanetary disk 4.56 Ga. They are partially composed of high-temperature components (refractory inclusions, chondrules, and Fe-Ni metal nuggets) formed by various nebular and/or planetary processes. These components are surrounded by complex, volatile-rich, fine-grained matrix material hosting the two main chondritic hydrogen-bearing phases: fluid-derived phyllosilicates and both soluble and insoluble organic matter (Le Guillou and Brearley, 2014). Matrix modal abundances are highly variable among chondrites, ranging from <5% in CB/CH chondrites to >95% in CI chondrites (Scott and Krot, 2014). Although the fine-grained matrices of different types of chondrites contain variable amounts of fluid-derived phyllosilicates and organic matter implying that their parent bodies accreted different amounts of water-ice grains and organic materials (Alexander et al., 2012; Vacher et al., 2020), matrix-rich chondrites (i.e., CI and CM carbonaceous chondrites) are enriched in H and C compared to their matrix-poor counterparts (i.e., ordinary and enstatite chondrites; Alexander et al., 2012; Vacher et al., 2020; Piani et al., 2020).

The source of the original water-ice grains and organics accreted by chondritic parent bodies is a long-standing debate. Hydrogen isotopes (expressed as D/H ratios) are powerful tracers of their origins because water-ice and organic grains inherited from the molecular cloud should be enriched in deuterium by ~2–3 orders of magnitude relative to those formed during the evolution of the protoplanetary disk (Ceccarelli et al., 2014; Cleeves et al., 2014; 2016). In the solar system, the H isotopic composition of water in planetary objects is generally thought to increase from very low values near the Sun (D/H ≈ $20 \times 10^{-6}$; Geiss and Gloecker, 2003) to intermediate D/H ratios in the inner solar system (e.g., D/H ≈ $150 \times 10^{-6}$ in Earth's ocean and bulk carbonaceous chondrites; e.g. Alexander et al., 2012; Vacher et al., 2020) and high D-enrichments in the outer solar system (e.g., up to $530 \times 10^{-6}$ in comet 67P/Churyumov-Gerasimenko; Altwegg et al., 2015). Nonetheless, direct comparisons between different planetary objects are not obvious because bulk chondrites correspond to complex mixtures of organics and hydrated minerals that cannot be mechanically separated for measurement of their specific D/H compositions. Consequently, chondrite bulk D/H ratios do not correspond to those of water-ice grains accreted by their respective parent bodies. In addition, the D/H ratios of cometary water are estimated from remote sensing measurements of deuterated vs. non-deuterated water molecules ($HDO/H_2O$) or fragments (e.g., OD/OH) in the vapor sublimated from comets (Bockelée-Morvan et al., 2015 and references therein), which cannot account for inputs from organic components and may be affected by sublimation-induced isotopic fractionations. Therefore, caution should be exercised when comparing the D/H ratios of different solar system objects and drawing conclusions on the origin(s) of water accreted by asteroidal and cometary bodies.

To better understand the process(es) responsible for the hydrogen isotopic variations observed among chondrites, we determined the D/H ratios of water in different types of carbonaceous chondrites (CI-, CM-, CO-, CR-type and Ungrouped) via *in-situ* secondary ion mass spectrometry (SIMS) measurements of the C/H and D/H ratios of chondritic matrices (Piani et al., 2018; Piani and Marrocchi, 2018). This technique allows the D/H ratios of hydrous minerals, and thus of initial water-ice grains accreted by chondritic parent bodies, to be determined without hydrogen contamination from adjacent organic grains (Piani et al., 2018). By comparing our isotopic results to literature data on water and organics, we propose a protoplanetary disk evolution scenario accounting for the D/H ratios observed in all chondrite groups. Our model reveals that the infall of molecular-cloud material (i) significantly contributed to chondritic water and organic budgets and (ii) occurred episodically for several millions of years, likely through filaments connecting the protosolar molecular cloud to the protoplanetary disk.

## 2. Material and methods

SIMS hydrogen isotopic measurements were performed on five CM-type (each with different degrees of aqueous alteration: Aguas Zarcas CM2.2, Martin and Lee, 2020; Jbilet Winselwan CM2.4/2.7, King et al., 2019b; Lonewolf Nunataks, LON 94101 CM2.2/2.3, Lindgren et al., 2011; Maribo CM2.6/2.7, van Kooten et al., 2018; and Mukundpura CM2.0, Rudraswami et al., 2019), one CR-type (Renazzo), one CO-type (Dominion Range, DOM 08006), two CI-type (Alais, Orgueil), one CY-type (Yamato, Y 980115) and three ungrouped (Bells, Essebi, and Tagish Lake) carbonaceous chondrites. Sub-millimeter pieces of each chondrite were handpicked under a stereomicroscope and pressed in pure indium. The Bells, Essebi, and DOM 08006 chondrites were taken from powders prepared for previous studies (Alexander et al., 2012, 2013; Howard et al., 2011, 2015). The Bells powder was made from two different pieces: samples C, from a stone recovered shortly after the fall, and W, from a stone recovered after a hurricane (Alexander et al., 2012). The pressed samples were not polished to avoid contamination and/or removal of soluble organics by the use of solvents. The workable nature of the fine-grained matrix material allows us to obtain flat areas of several hundreds of micrometers. The matrix areas were then identified using a polarized reflected-light microscope and backscattered electron images and chemical maps acquired by dispersive X-ray spectroscopy (EDS). Chondrule fragments, areas with large silicates or holes were avoided based on EDS maps. A final verification was performed after SIMS measurements to verify the position of the SIMS analyses. Analyses with positions that did not correspond to flat fine-grained matrix were removed from the final dataset. The samples were stored in a vacuum cabinet before and after analyses and were introduced in the SIMS instruments several days before measurement.

Following our previous works (Piani et al., 2018; Piani and Marrocchi, 2018), a series of reference materials including hydrated minerals (montmorillonite and serpentine), hydrogen-bearing glasses, and terrestrial and extraterrestrial D-rich organic matter were used to calibrate the SIMS and correct for instrumental mass fractionation on the D/H ratio. All samples and standards were pressed in indium and gold-coated before analysis.

All SIMS analyses were performed using the CAMECA IMS-1280HR2 instruments installed at Hokkaido University (Japan) and the CRPG-CNRS (France) during five analytical sessions between December 2015 and July 2019. The analytical conditions used are detailed in Piani et al. (2018) and Piani and Marrocchi (2018) and briefly summarized here. A 10 keV $Cs^+$ primary beam was used for the measurements. The vacuum in the analytical chamber was always below $5 \times 10^{-9}$ mbar. Prior to analyses, the samples were pre-sputtered at high current (1.5–2.5 nA) for 5–8 minutes to clean the sample surface, remove the adsorbed H and reach the sputtering steady-state. The samples were then sputtered with a 80–500 pA beam either (1) shaped by an aperture in the primary column, allowing a large homogeneous ellipsoidal shape with a major axis of about 70 μm and a minor axis of about 50 μm, or (2) rastered over a $50 \times 50$ μm$^2$ area. A high-magnification mode (Max Area 40) and a small field aperture (FA 2000) were used to minimize H contamination that would diffuse from the border of the beam. A normal-incidence electron gun was used for charge compensation. In the case of the shaped beam, the analyzed area was restricted to a $10 \times 10$ μm$^2$ area in the center of the ellipse by using the small field aperture and high-magnification mode to remove H coming from the border of the analyzed area. In the case of the rastered beam, the analyzed area was additionally restricted to a $15 \times 15$ μm$^2$ area in the center of the rastered area by using a 30% electronic gate to remove H coming from the border. Identical results were obtained for the standards and some reference samples with the two beam settings (Piani et al., 2018; Piani and Marrocchi, 2018). $H^-$, $D^-$, $^{13}C^-$, and $^{29}Si^-$ ions were collected successively by changing the magnetic field and counted with the monocollection electron multiplier. The mass resolution power was set to $M/\Delta M$ = 3,300 to avoid interferences on $^{13}C^-$ by $^{12}CH^-$ and on $^{29}Si^-$ by $^{28}SiH^-$. For each analysis, 30–50 cycles were collected with 1 s of counting time per cycle for $H^-$, $^{13}C^-$, and $^{29}Si^-$ and 10–20 s per cycle for $D^-$, totaling 30 minutes per analyses. Due to the lower abundance of $^{13}C$ compared to $^{12}C$, we were able to measure carbon with the electron multiplier across the entire range of C concentrations, from the C-poorest to the C-richest matrix areas. No clear relation between $^{29}Si^-$ and $^{13}C^-$, $H^-$, or $D^-/H^-$ was observed in the matrices. The statistical error on D/H in the sample having the lowest D/H ratio was 3 % (2σ) and the reproducibility on the reference materials was ≤ 8 % (2σ standard deviation). The statistical error for the $^{13}C/H$ ratio on the samples with low C-content is typically of 6 % (2σ). The D/H ratios estimated for water in chondrites were corrected for instrumental mass fractionation using a calibration line determined using the standard materials (Remusat et al., 2016; calibration parameters in Table 1).

The zero-intercept of the $D^-/H^-$ vs. $C^-/H^-$ correlation in each chondrite's matrix was used to estimate the D/H ratio of water in that chondrite (Table 1; see Piani et al., 2018 for details). Because the main H-bearing phases in carbonaceous chondrite matrices correspond to hydrated silicates (Fe-Mg serpentine, saponite, and cronstedtite), and because the equilibrium isotopic fractionation factor α between hydrated silicates and water is lower than the D/H reproducibility on the reference materials, we consider the zero-intercept to be a direct proxy for the D/H ratio of the water from which the minerals formed (Piani et al., 2018; Piani and Marrocchi, 2018).

## 3. Results

From our SIMS data measured in the chondrite matrices, we obtained correlations between D/H and C/H for ten of the twelve analyzed chondrites (Pearson coefficients from 0.33 to 0.78; Table 1; Figs. 1 and S1). No correlation was observed for the chondrites Tagish Lake (ungrouped) and Y-980115 (CY; Table 1; Figs. 1b and S1). For the ten others, the D/H ratio of water was estimated from the zero-intercept and corrected for instrumental mass fractionation using the standard calibration measured during the same session (Table 1, Fig. 1).

Water in four of the five CM-type carbonaceous chondrites has D/H ratios that are identical within error (2σ, Table 1): Aguas Zarcas ($107 \pm 41 \times 10^{-6}$), LON 94101 ($94 \pm 24 \times 10^{-6}$), Maribo ($90 \pm 27 \times 10^{-6}$), and Mukundpura ($105 \pm 18 \times 10^{-6}$). As observed by Piani et al. (2018), the degree of aqueous alteration of a chondrite does not obviously affect the D/H ratio of its water. Compared to the other CM chondrites, only water in Jbilet Winselwan (CM) has a distinctly higher D/H ratio ($162 \pm 87 \times 10^{-6}$) and a lower regression coefficient (Pearson coefficient of 0.33) indicating greater scattering of the data around the regression line (Table 1). Water in the two CI chondrites Alais and Orgueil has consistent D/H values of $170 \pm 39 \times 10^{-6}$ and $172 \pm 42 \times 10^{-6}$, respectively (Table 1). Water in the CO chondrite DOM 08006 is estimated to have a D/H ratio of $203 \pm 21 \times 10^{-6}$ (Table 1), although the correlation in this sample shows a low Pearson correlation coefficient of 0.33 (Table 1). Water in the two ungrouped chondrites Essebi and Bells has D/H ratios of $220 \pm 46 \times 10^{-6}$ and $210 \pm 19 \times 10^{-6}$, respectively (Table 1; Fig. 1), and is thus enriched in deuterium compared to water in the CI, CM, and CO chondrites. In contrast to the Bells C sample (D/H = $210 \pm 19 \times 10^{-6}$), water in the Bells W sample presents a lower D/H value ($167 \pm 22 \times 10^{-6}$) that might indicate either terrestrial contamination of this late-recovered sample or heterogeneity within the Bells chondrite. Water in the CR chondrite Renazzo has a D/H ratio of $360 \pm 32 \times 10^{-6}$ (Table 1), the most deuterium-rich value yet obtained for water in carbonaceous chondrite matrices.

## 4. Discussion

### 4.1 Hydrogen isotopic compositions of chondritic water

The D/H ratios determined in this study for water in CM, CI, CO, and CR chondrites span a large range of values, from 90 to 360 × $10^{-6}$ (Fig. 2, Table 1). In Figure 2, these values are reported alongside those estimated in our previous works for water in CV and other CM chondrites (Piani et al., 2018; Piani and Marrocchi, 2018) and by bulk analyses of CM and CR chondrites (Alexander et al., 2012). Except for Jbilet Winselwan, our D/H ratios estimated for water in CM chondrites (90–107 × $10^{-6}$) are similar within errors to the mean value obtained previously using the same technique (101 ± 6 × $10^{-6}$, Fig. 2; Piani et al., 2018). Excluding the peculiar CM chondrites Jbilet Winselwan and the D-rich Paris (Vacher et al., 2016; Piani et al., 2018), the average D/H ratio of water in all CMs analyzed is 100 ± 4 × $10^{-6}$ (2σ). This value is in agreement with those inferred from hydrated minerals in Maribo (103–114 × $10^{-6}$; van Kooten et al., 2018), but is slightly higher than that determined from the bulk measurements of 45 CM chondrites characterized by different degrees of aqueous alteration (87 ± 8 × $10^{-6}$, 2σ, Fig. 2; Alexander et al., 2012). Our D/H ratio for water in the CR chondrite Renazzo (360 ± 32 × $10^{-6}$, 2σ) is significantly higher than that estimated from bulk measurements of 11 CR carbonaceous chondrites ($96^{+110}_{-65}$ × $10^{-6}$, 2σ; Alexander et al., 2012). However, we consider that their D/H value for water is more representative than our estimate based on a single meteorite. Because only a single CO chondrite was measured in this study, our estimate might not be representative of water in other CO chondrites and will be omitted from further discussion.

An important outcome of our results is that the average D/H ratios of water are distinct among the different chondrite groups: $149 \pm 10 \times 10^{-6}$ ($2\sigma$), $100 \pm 4 \times 10^{-6}$ ($2\sigma$), $171^{+17}_{-10} \times 10^{-6}$ ($1\sigma$; Alexander et al., 2012), $171 \pm 16 \times 10^{-6}$ ($2\sigma$), and $215 \pm 10 \times 10^{-6}$ ($2\sigma$) in CV, CM, CR, CI, and ungrouped chondrites, respectively (Fig. 2). These values are either similar to (CV, CI, and ungrouped) or lower than (CM and CR) those obtained for the bulk chondrites (Fig. 3; Table S1 and references therein), likely reflecting the accretion of variable amounts of organic matter with varying D enrichments. The range of water D/H values observed in carbonaceous chondrites (CCs) is thus small and D-depleted compared to the most pristine ordinary chondrite (OC) Semarkona (D/H = $393–609 \times 10^{-6}$; Deloule and Robert, 1995; Piani et al., 2015) and cometary water (D/H = $140–650 \times 10^{-6}$; Altwegg et al., 2017 and references therein; Fig. 3 and 4).

The D/H ratios of the main component of chondritic organic matter, insoluble organic matter (IOM), have been measured in a large set of samples (Alexander et al., 2007) and show a wider range than those of water in both CCs and non-carbonaceous chondrites (NCs; Fig. 4). IOM D/H ratios in the three groups of OCs ($870 \times 10^{-6}$, $660 \times 10^{-6}$, and $710 \times 10^{-6}$ in H, L, and LL groups, respectively; Table S1) are about 4 times the average value of IOM in enstatite chondrites (D/H = $153 \times 10^{-6}$; Table S1, Fig. 4; Alexander et al., 2007; Piani et al., 2012). In carbonaceous chondrites, IOM D/H ratios increase from CV/CO (~$220 \times 10^{-6}$) to CM/CI (~$300 \times 10^{-6}$) and finally to CR/ungrouped (>$600 \times 10^{-6}$; Table S1; Fig. 4).

The origin of these large hydrogen isotopic variations in chondritic water and IOM in both NCs and CCs remains unclear and controversial. These variations have been attributed to several processes and/or locations within the protoplanetary disk, including early fluid circulation during the evolution of chondritic parent bodies, physico-chemical processes operating in the solar protoplanetary disk, and inheritance from the molecular cloud. In the following section, we discuss these different possibilities in light of our new results and using diagrams compiling the D/H values of chondritic water, IOM, and the bulk meteorites (Fig. 3 and 4).

**4.2 Origin of hydrogen isotopic variations in chondritic water and organics**

*4.2.1 Role of parent-body processes*

It has been proposed that secondary parent-body processes could be at the origin of the hydrogen isotopic variations observed in chondrites due to (1) the aqueous oxidation of iron-bearing phases followed by $H_2$ loss and Rayleigh-type isotopic fractionation (Alexander et al., 2010; Sutton et al., 2017) and/or (2) hydrogen isotopic exchange between organic components and aqueous fluids (Alexander et al., 2010; Bonal et al., 2013). The former process is considered to be most effective for enriching the most water-depleted chondrites (i.e., OCs, CV and CO-type CCs) in deuterium because their bulk D/H ratios are not buffered by the large amount of remaining water (Alexander et al., 2010). The efficiency of the latter process should depend on the temperature and duration of the interaction between water and organics as well as the nature of the organics and their ability to exchange hydrogen with water (e.g., Sessions et al., 2004). However, no experimental study has measured the putative H isotopic fractionation produced through the anaerobic corrosion of metal, and tests of water-organic isotopic exchanges under asteroidal conditions revealed complex patterns (Foustoukos et al., 2021). In addition, these models are based on the strong and debatable assumptions that (i) all chondrites accreted water-ice and organic matter with similar respective D/H ratios and (ii) asteroidal parent-body processes have swamped any D enrichments inherited from solar-system or molecular-cloud processes (Alexander et al., 2010; Sutton et al., 2017). Consequently, the parent-body model remains speculative and difficult to comprehend given the complexity of organic components in meteorites (De Gregorio et al., 2010; Remusat et al., 2010). Based on our data and a comprehensive hydrogen isotopic database, we show in the following paragraphs that secondary parent-body processes, although important, cannot be advocated as the main processes controlling the observed D/H variations of chondrites and their constituents.

*Ordinary chondrites.* Highly heterogeneous D/H values and extreme D-enrichments up to D/H = $1,800 \times 10^{-6}$ have been reported in the matrix of the least altered OC LL3.00 Semarkona (Deloule and Robert, 1995; Grossman and Brearley, 2005; Piani et al., 2015). Although OCs display the most D-rich bulk values among chondrites, their IOM is less enriched in D and more homogenous (showing only rare, micron-sized D-rich anomalies, or "hotspots"; Remusat et al., 2016) than some of the extremely heterogeneous hydrated minerals found in Semarkona (Piani et al., 2015). These facts do not support OM-water isotopic exchanges following the D enrichment of water through Rayleigh-type fractionations (Alexander et al., 2010). Moreover, if IOM was enriched in D by isotopic exchange with water that was itself enriched in D due to $H_2$ loss, the bulk D/H ratios of moderately metamorphosed OCs should be enriched in D compared to the least metamorphosed OCs. However, D enrichments are observed even in the most pristine OCs (i.e., <3.2; Marrocchi et al., 2020) whereas mildly to highly metamorphosed chondrites show D-poor bulk compositions (McCubbin and Barnes, 2019; Vacher et al., 2020). This strongly suggests that the D enrichments and heterogeneities observed at both mineral and bulk scales in OCs do not result from $H_2$ loss during parent-body processes but were inherited from ice precursors.

*CM chondrites.* Excluding Paris and Jbilet Winselwan, water in all CM chondrites displays similar D/H ratios, regardless of the degree of alteration (Fig. 2; Piani et al., 2018).

The Paris chondrite is a specific case because water in the least altered lithology is significantly enriched in D compared to the more altered lithology, which shows D/H values close to the average for CM water (Piani et al., 2018). Because Fe-Ni metal beads have been highly oxidized in the altered Paris lithology but remain unaltered in the most pristine regions (Hewins et al., 2014; Marrocchi et al., 2014), the opposite isotopic characteristics are expected if Rayleigh-type fractionation following metal oxidation and $H_2$ loss were the dominant processes controlling D/H variations. Furthermore, the similarity of the water D/H ratios estimated by SIMS ($100 \pm 4 \times 10^{-6}$; Piani et al., 2018) and bulk measurements ($87 \pm 8 \times 10^{-6}$; Alexander et al., 2012) implies that isotopic re-equilibration between water and organics was extremely limited during the evolution of the CM parent body(ies). Indeed, any isotopic exchange would have increased the D/H ratio of water while decreasing that of organics, the intensity of isotopic variations being dependent on their relative proportions (e.g., Foustoukos et al., 2021). If that were the case, important isotopic differences would be particularly expected for chondrites characterized by varying degrees of alteration, such Cold Bokkeveld (CM2.2, highly altered), LON 94101 (CM2.2/2.3, mildly altered), and Murray (CM2.5, poorly altered), which is not the case (Fig. 2 and Fig. 5; Piani et al., 2018). The lack of IOM-water isotopic equilibration in CMs implies the existence of an additional C-bearing component having higher D/H values than IOM. Importantly, recent bulk hydrogen measurements performed on a series of CCs and OCs revealed the importance of removing atmospheric contamination when measuring H contents and isotopic compositions (Vacher et al., 2020). They performed pre-degassing at 120 °C during 48 h under vacuum prior to H measurements to remove atmospheric water adsorbed on the sample surface. The lower slope obtained in the D/H vs. C/H plot for samples for which the atmospheric contamination was reduced (Fig. 6A) is inconsistent with the IOM of CMs being initially as D-rich as the IOM in CRs (Fig. 6B), indicating that they inherited isotopically distinct IOM in their respective parent bodies.

*CI, CR, and Bells/Essebi chondrites.* These chondrites are characterized by (i) D-rich water and IOM compared to CMs (Fig. 4) and (ii) the presence of D-rich hotspots in IOM (e.g., Busemann et al., 2006). In addition, large D/H variations have been reported in hydrated minerals in CR chondrites (Bonal et al., 2013). These characteristics imply that water-IOM isotopic exchanges cannot be at the origin of the H isotopic differences between these meteorites and CMs. In addition, Rayleigh-type fractionation is not expected to be efficient in water-rich meteorites (Sutton et al., 2017). Furthermore, small-scale heterogeneities in IOM and hydrated minerals are unlikely to result from parent-body processes unless invoking unconstrained, localized kinetic reactions.

*CV chondrites.* The D/H ratios of water and IOM in oxidized CVs are D-rich and D-poor, respectively, compared to CM water and IOM (Alexander et al., 2007; Piani and Marrocchi, 2018). These peculiar characteristics could result from isotopic exchange during parent-body alteration, assuming an initial CM-like H isotopic composition for water and organics (Piani and Marrocchi, 2018). However, they could also result from the accretion of water-ice and organics from different reservoirs characterized by variable H isotopic compositions (Piani and Marrocchi, 2018).

*Jbilet Winselwan (CM), Tagish Lake (C-ungrouped), and Y-980115 (CY).* The absence of a positive correlation between D/H and C/H in these peculiar chondrites (Table 1) may suggest perturbations of the initial isotopic signatures of their organic and water components. Such perturbations might have resulted from impact-induced dehydration in Jbilet Winselwan (King et al., 2019a), thermal metamorphism (>500 °C, King et al., 2019b) and/or terrestrial weathering in Y-980115 (e.g., Alexander et al., 2018), or IOM-water isotopic exchange and/or the possible loss of a D-rich component in Tagish Lake (Herd et al., 2011). Another possible explanation for Tagish Lake could be that the abundant carbonates, that can represent half of the total carbon content (Grady et al., 2002), could influence the measured C/H ratio independently to the D/H ratio.

Thus, the H isotopic compositions of bulk CCs and their water and IOM do not support a preponderant role of parent-body processes in controlling D/H variations among chondrites. CV chondrites are the only group in which isotopic exchanges between water and organics remain possible or even favorable due to the high hydrothermal temperatures of the CV parent body (Ganino and Libourel, 2017). In addition, only a few specific meteorites (Jbilet Winselwan, Tagish Lake, Y-980115) experienced hydrothermal alteration and impact heating sufficient to modify their primordial isotopic signatures. These results thus suggest that the H isotopic features of chondrites are mostly inherited from processes that operated within the solar system and/or the parent molecular cloud.

*4.2.2 Fingerprints of the protosolar molecular cloud in meteoritic water and organics*

Astronomical observations have revealed that water molecules in the interstellar medium (ISM) are significantly enriched in deuterium relative to molecular hydrogen, with D/H ratios of water in protostars reaching up to a few $10^{-2}$ (e.g., Ceccarelli et al., 2014 and references therein). Similarly, highly D-rich organics (i.e., $2–7 \times 10^{-2}$; Roberts et al., 2002) are produced in dense regions of the ISM due to the cold temperatures and high levels of galactic cosmic radiation (Geiss and Gloeckler 2003). Protoplanetary disks are by-products of star formation, resulting from the conservation of angular momentum of the parent molecular cloud from which they formed. Observational and theoretical evidence show that protoplanetary disks are initially compact, with most infalling material from the molecular cloud being injected close to the protostars (Fig. 7; Pignatale et al., 2018; Zhao et al., 2020). While being fed by the molecular cloud, the disks expand through viscous spreading (Hueso and Guillot, 2005) and their centrifugal radii, the distance within which the parental cloud material is injected, increase with time.

Infalling D-rich interstellar water-ice grains will experience sublimation upon injection into the inner tens of astronomical units of protoplanetary disks (Visser et al., 2009). Subsequently, the D/H ratio of water will decrease toward bulk chondrite values (i.e., ~150 × $10^{-6}$) through isotopic exchange with the D-poor protosolar $H_2$ (Jacquet and Robert, 2013; Yang et al., 2013) via the reaction:

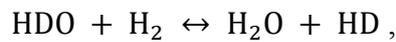
$$HDO + H_2 \leftrightarrow H_2O + HD,$$

whose equilibrium fractionation factor $\alpha_{H_2O-H_2}$ is temperature-dependent, increasing with decreasing temperature (Richet et al., 1977). Later ionization-driven chemical reactions during the disk's evolution do not efficiently enrich water in deuterium (Cleeves et al., 2014; Roskosz et al., 2016). This implies that the range of water D/H ratios observed among chondrites (i.e., 90–500 × $10^{-6}$; Figs. 3 and 4) corresponds to a mixture between (i) water formed by sublimation, high-temperature isotopic equilibration, and/or recondensation during the evolution of the disk and (ii) inherited D-enriched interstellar ices unaffected by disk processes (Jacquet and Robert, 2013; Yang et al., 2013). The inheritance of molecular cloud material may become even more obvious when considering organic grains because, compared to chondritic water, they show higher and more variable D/H (e.g., Busemann et al., 2006; Alexander et al., 2007; 2012) that cannot be reproduced by disk-ionization models (Cleeves et al., 2016). Nonetheless, organo-synthesis triggered by organic radicals produced in ionized disk environments may at least partly explain the D/H heterogeneities observed in meteoritic organics (Robert et al., 2017).

Based on the above observations, we propose a model in which incoming water-ice and organic grains from the molecular cloud episodically fuel the solar protoplanetary disk throughout its evolution. The lifetimes of star-forming molecular clouds are estimated to be ~20–30 Myr (Murray, 2011), implying that D-rich water-ice and organic grains from the ISM can be injected into protoplanetary disks throughout their evolution and until their dissipation (i.e., 5–10 Myr). Water-ice and organic grains have different vaporization temperatures (the so-called snow and tar lines; Figs. 7 and 8), with refractory organic grains being stable at much higher temperatures (i.e., above 350–450 K) than water-ice grains (i.e., ~160 K, Fig. 7; Nakano et al., 2003; Kuramoto and Yurimoto, 2005; Bermingham et al., 2020). Consequently, the tar line is inward of the snow line, defining three regions in the disk (#1–3, Fig. 8) where (#1) refractory organic and water-ice grains are sublimated, (#2) refractory organic grains are stable but water is in vapor form, and (#3) both refractory organic and water-ice grains are unaffected. As the disk temperature drops, region #2 will grow due to (i) the inward drift of the tar line and (ii) the fossilization of the snow line at 3 astronomical units (Morbidelli et al., 2016). Consequently, within region #2, water-ice grains experience continued sublimation and isotopic exchange with D-poor $H_2$ whereas refractory organics do not, leading to an isotopic decoupling (Fig. 4). This implies that most organic grains are not affected by thermal reprocessing in the disk and maintain their D enrichments and D/H variability. Moreover, the synthesis of organic matter with D-poor signatures in ionized disk environments (e.g., Bekaert et al., 2018) and/or under aqueous conditions on asteroidal parent bodies (e.g., Kebukawa et al., 2013) may result in a final mixture of chondritic organic compounds bearing large H isotopic variations. In contrast to refractory organics, water experiences isotopic exchange with the gas until the disk temperature becomes cold enough for isotopic exchange reactions and sublimation to be ineffective at ~200 K and 160 K, respectively (Fig. 7; Jacquet and Robert, 2013). This naturally explains why hydrated chondritic minerals (and thus water) are less enriched in deuterium and less isotopically variable than organics (Fig. 4; Alexander et al., 2010, 2012; Piani et al., 2015). Thus, depending on the timing of accretion of chondritic parent bodies, water and organics may be characterized by drastically different D/H ratios.

Accordingly, in this model, several scenarios are possible:

1) Chondrites that accreted early in the disk's history are characterized by D-poor water and organics (e.g., enstatite chondrites, COs, and CVs; Fig. 4).

2) Chondrites that accreted below the organic sublimation temperature comprise D-rich organics, but water variably depleted in deuterium compared to the organics (e.g., CMs, CIs, and C-ungrouped; Fig. 4). The D-depleted water in CM chondrites could thus be the result of a more efficient isotopic re-equilibration with $H_2$, possibly at higher temperatures and/or during longer interactions than for CI and C-ungrouped chondrites.

3) Finally, chondrites that accreted when the disk cooled to <200 K comprise both D-rich water and organic grains (e.g., OCs and CRs; Fig. 4).

This model implies that chondrites have continuously recorded processes (i.e., sublimation and isotopic exchange) that eventually became inefficient during the evolution of the protoplanetary disk, thus allowing for the highly heterogeneous isotopic compositions of their constituent hydrated minerals (Bonal et al., 2013; Piani et al., 2015) and/or organics (Busemann et al., 2006).

A key question regarding chondrites concerns their accretion ages. As the background temperature of the disk controlled the efficiency of sublimation and isotopic exchange processes (e.g., Jacquet and Robert, 2013), our model suggests that the H-isotopic compositions of both water and organics are also directly related to the timing of accretion of chondrites. It also suggests that the isotopic decoupling between water-ice grains and IOM may have occurred asynchronously between the NC and CC reservoirs because their respective thermal evolutions likely differed, in line with recent models (e.g., Lichtenberg et al., 2021). Of note, our qualitative model is coherent with estimates of the timing of accretion of chondritic parent bodies, with (i) enstatite chondrites having accreted before OCs and CCs and (ii) among CCs, water-poor CO, CV, and CK chondrites having accreted before water-rich CR, CM, and CI chondrites (Sugiura and Fujiya, 2014; Desch et al., 2018). Sugiura and Fujiya (2014) also suggest that OCs, which formed in the inner regions of the disk, may have been accreted before CCs, making the large H isotopic variabilities observed in OCs counterintuitive, as one would expect that the inner disk remained hotter for a longer duration than the outer disk. However, recent estimates of the ages of OC chondrules using Al-Mg and Pb-Pb systematics have shown that they formed up to 3 Myr after the formation of Ca- and Al-rich inclusions (Bollard et al., 2017; Pape et al., 2019). Because chondrules formed before the accretion of planetesimals, this implies that the OC and CC parent body(ies) formed contemporaneously, further highlighting the isotopic decoupling between water-ice grains and IOM (Fig. 4).

Because some D-rich organics and waters are not at isotopic equilibrium with the gas of the disk (i.e., largely different water and IOM values per chondrite class in Fig. 4), our data suggest that they did not experience the high-temperature chondrule-forming events that would have induced sublimation, isotopic exchange, and/or recondensation. Hence, we propose that the isotopically heterogeneous D-rich ices and organics observed in OCs and CRs correspond to materials that arrived and were reprocessed late in the history of the disk. Interestingly, this suggests that both the inner and outer parts of the disk were, at some point, cold enough for unprocessed, D-rich interstellar water-ice and organic grains to accrete (Fig. 7). This also reiterates that the parent molecular cloud episodically fed the disk for at least several million years, and possibly throughout the disk's history, through filaments connecting both structures (Hennebelle et al., 2016; Zhao et al., 2020). Such structures −observed at all scales in the interstellar medium, from molecular clouds to individual stars− result from prominent contracting forces induced by gravity, turbulences and/or magnetic fields (e.g., Federrath, 2016). They play a key role in the star-forming processes and could have contributed to the establishment of the peculiar H isotopic signature of our solar system.

## 5. Concluding remarks

We determined the hydrogen isotopic compositions of water in a large set of chondritic samples (CI, CM, CO, CR, CY, and C-ungrouped) by *in-situ* SIMS analyses. Our main result is that the D/H ratios of chondritic water appear to be distinct and unique among the different groups of carbonaceous chondrites. This result can be used for the characterization and comparison of the samples returned from the asteroids Ryugu and Bennu by the spatial missions Hayabusa 2 (JAXA; 5.4 g of samples delivered to Earth in December 2020) and OSIRIS-REx (NASA; to be returned on Earth in 2023). Waters in CCs are D-depleted and show a restricted range of D/H values relative to the most pristine ordinary chondrite and cometary water.

From these results, and by comparison with literature data on chondritic water and organics, we drew the following conclusions.

(i) The D/H ratios of water and IOM do not support a preponderant role of parent-body processes in controlling the D/H variations of carbonaceous and ordinary chondrites.

(ii) D/H variations among chondrites represent a mixture between water formed by sublimation, high-temperature isotopic equilibration, and/or recondensation during the evolution of the disk and inherited D-rich interstellar ices unaffected by those processes.

(iii) Based on their different sublimation temperatures, water-ice experienced sublimation and isotopic exchange with molecular $H_2$ for a longer duration than refractory organics. This induced an isotopic decoupling through which organics preserve more D-rich signatures than water.

(iv) Hydrated minerals and/or organics in later-accreted chondrites may record extreme and highly heterogeneous D/H values inherited from the interstellar medium.

(v) The molecular cloud episodically fed the protoplanetary disk for several millions of years through filaments connecting both structures.

**Appendix A. Supplementary material**

Table S1 presents the literature data for bulk, IOM and water D/H ratios used in Fig. 3 and 4. Fig. S1 presents the measured D/H vs. $^{13}$C/H ratios measured by SIMS in the matrices of all carbonaceous chondrites analyzed in the present study (Table 1). Linear fits (solid lines) and 95% confidence interval bands (dashed lines) are shown for all chondrites. Error bars represent 2σ internal errors.

**Appendix B. Supplementary material**

SIMS data related to this article can be found on-line at https://doi.org/10.24396/ORDAR-61.

**Acknowledgments**

Conel Alexander is warmly thanked for stimulating discussions on the onset of parent-body processes. Patrick Hennebelle and Sébastien Charnoz are warmly thanked for intense and fruitful discussions on the interactions between molecular clouds and accretion disks. We thank Conel Alexander and Kieren Howard for providing the Bells, Essebi, and DOM 08006 powders. We also thank the National Institute of Polar Research (Japan), the Natural History Museum of Denmark (Copenhagen), and the Museum national d'Histoire Naturelle (Paris, France) for loaning samples. This work was funded by l'Agence Nationale de la Recherche through grant ANR-19-CE31-0027-01 HYDRaTE (PI Laurette Piani). This is CRPG contribution #2728.

**Figure caption**

**Fig. 1.** Measured D/H vs. $^{13}$C/H ratios in the matrices of representative carbonaceous chondrites Mukundpura (CM), Jbilet Winselwan (CM), Aguas Zarcas (CM), Essebi (Ungrouped), Bells (pieces C and W; Ungrouped), Alais (CI), and Y-980115 (CY). The data were acquired during two analytical sessions in April 2018 (left) and July 2019 (right). Linear fits (solid lines) and 95% confidence interval bands (dashed lines) are shown for all chondrites except Y-980115, which does not show any correlation (see Table 1). Error bars represent 2σ internal errors.

**Fig. 2.** D/H ratios of water estimated from the D/H vs. C/H correlation measured by SIMS for all chondrites measured so far by SIMS. Error bars indicate the 95% error estimated from SIMS analyses of the samples and standards. Water D/H ratios estimated from bulk measurements of CM and CR chondrites (Alexander et al., 2012) are also reported with 2σ errors. Superscripts indicate data references: 1, Piani et al., 2018; 2, Piani and Marrocchi, 2018; 3, this study; 4, Alexander et al., 2012.

**Fig. 3.** Bulk and water D/H ratios measured in carbonaceous chondrites plotted as a function of the putative heliocentric distance at which their parent bodies formed (Desch et al., 2018). Bulk data are from a literature compilation (see Table S1). Water D/H ratios are from Alexander et al. (2012) for CR chondrites, Piani and Marrocchi (2018) for CV chondrites, Piani et al. (2018) and the present study for CM chondrites, and only the present study for CI and Ungrouped chondrites.

**Fig. 4.** Bulk, water, and insoluble organic matter (IOM) D/H ratios of non-carbonaceous (NC) and carbonaceous chondrites (CC). Bulk data are from a literature compilation (see Table S1) and IOM data from Alexander et al. (2007).

**Fig. 5.** Schematic view of expected (A, B) and measured (C) D/H ratios of water in CM chondrites. (A) Schematic view of the expected D/H ratio of CM water for the CM chondrites C1, C2, and C3 proposed from bulk measurements by Alexander et al. (2012), assuming isotopic exchange between water and organic matter. The initial D/H value of water would be enriched in D due to isotopic exchange with D-rich organic matter (gray arrows). Bells (ungrouped) is here considered to have behaved like a CM chondrite. (B) Based on bulk and IOM data (Alexander et al., 2007, 2012) and assuming that organics were in H isotopic exchange with water during parent-body alteration (Alexander et al., 2012), the D/H ratio of water in decreasingly altered CMs (Cold Bokkeveld < LON 94101 < Murray) is expected to increase from about $116 \times 10^{-6}$ to $144 \times 10^{-6}$. The water in Bells is expected to have D/H = $103 \times 10^{-6}$. (C) As reported in this study (Table 1; Fig. 2) and schematically represented here, such isotopic variations are not observed for CM water measured by SIMS: water in all CMs displays the same value ($100 \pm 4 \times 10^{-6}$) close to the initial D/H value of water estimated from bulk measurements (Alexander et al., 2012), whereas water in Bells water has a significantly higher D/H ratio of $210 \pm 19 \times 10^{-6}$ (Table 1).

**Fig. 6.** (A) Comparison of the bulk D/H and C/H ratios obtained for CM chondrites with (Vacher et al., 2020) and without (Alexander et al., 2012) pre-degassing of the samples (48 h, 120 °C) prior to analyzing their hydrogen isotopic compositions (from Vacher et al., 2020). Pre-degassing minimizes atmospheric contamination, resulting in the measurement of lower H contents in CM chondrites (Vacher et al., 2020). The linear regressions obtained for these two sets of data have distinct slopes but similar intercepts, within error. (B) Interestingly, the D/H ratios of IOM in CM chondrite are consistent with the extrapolation of the regression based on the pre-degassed measurements, but are inconsistent with those performed without pre-degassing.

**Fig. 7.** Schematic model of chondrite formation during the growth of the protoplanetary disk from the molecular cloud: (left) large-scale view, (center) view at the scale of the solar system, and (right) the background temperature as a function of time. (A) The young, compact disk receives material from the molecular cloud only in its most inner portions. At this time, the temperature is sufficiently high to sublimate both water-ice and organic grains inherited from the molecular cloud. (B, C) As the disk grows through viscous spreading, continued injections of molecular-cloud material occur at increasing centrifugal radii. At this time, the disk may have separated into inner and outer reservoirs, but both remained hot enough to sublimate water and organics. (D) As the disk cools, an isotopic decoupling occurs, in which water-ice grains continue to sublimate and experience isotopic exchange, whereas organics do not due to their higher sublimation temperature. This implies that organics may maintain D enrichments inherited from the interstellar medium, whereas water experienced continued isotopic exchange with molecular $H_2$ until the disk cooled enough for isotopic exchange to become inefficient.

**Fig. 8.** Schematic representation of the snow and tar lines. The snow line is the closest distance to the Sun at which ice condenses, and the tar line is that at which organic grains are stable. These lines define three regions where refractory organic and water-ice grains are sublimated (region #1), only refractory organic grains are stable whereas water exists only in vapor form (region #2), and both organic and water-ice grains are stable (region #3).

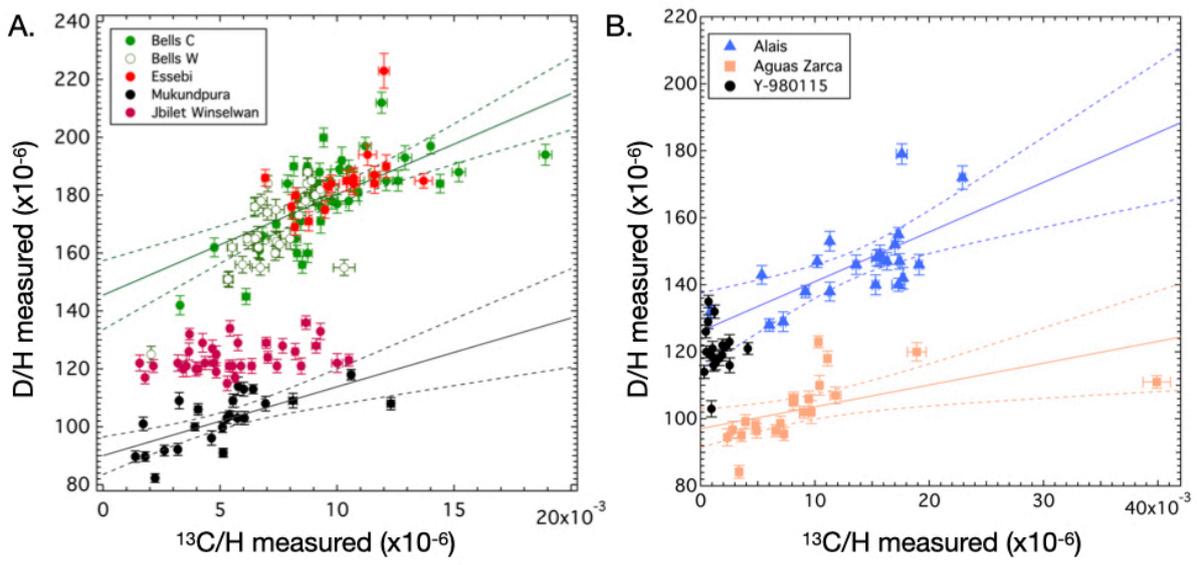

**Fig. 1**

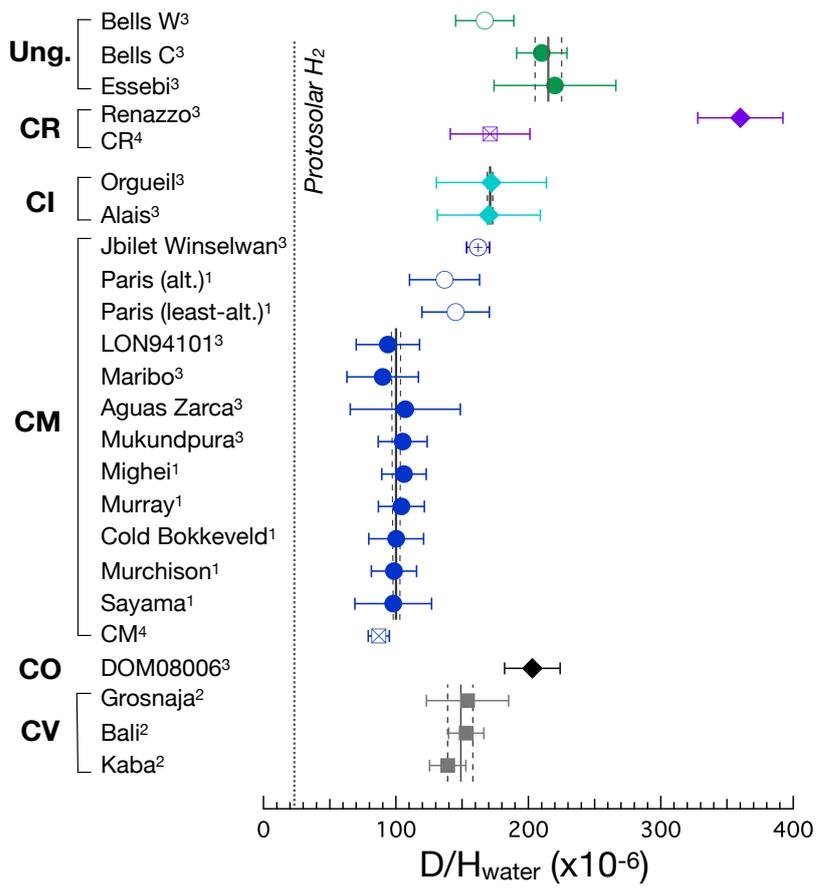

**Fig. 2**

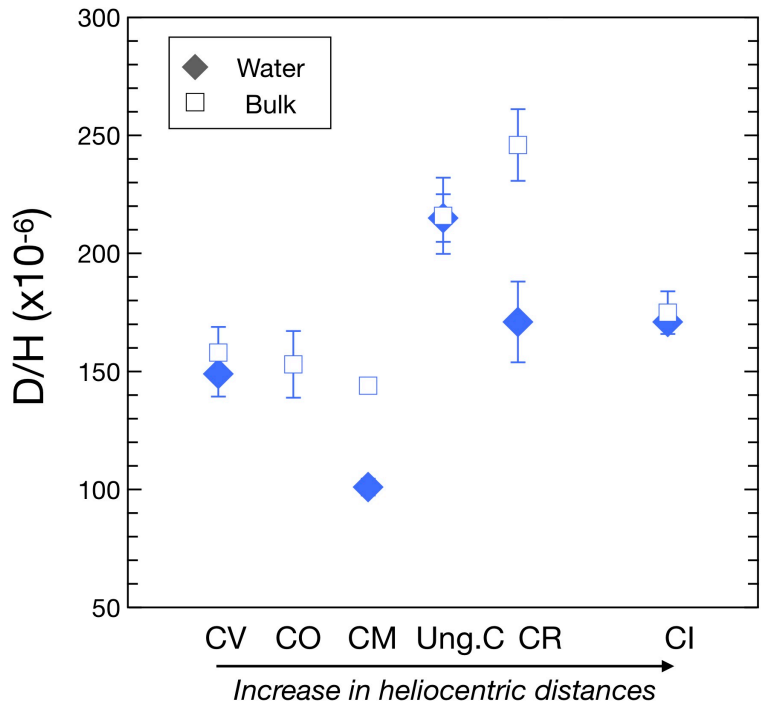

**Fig. 3**

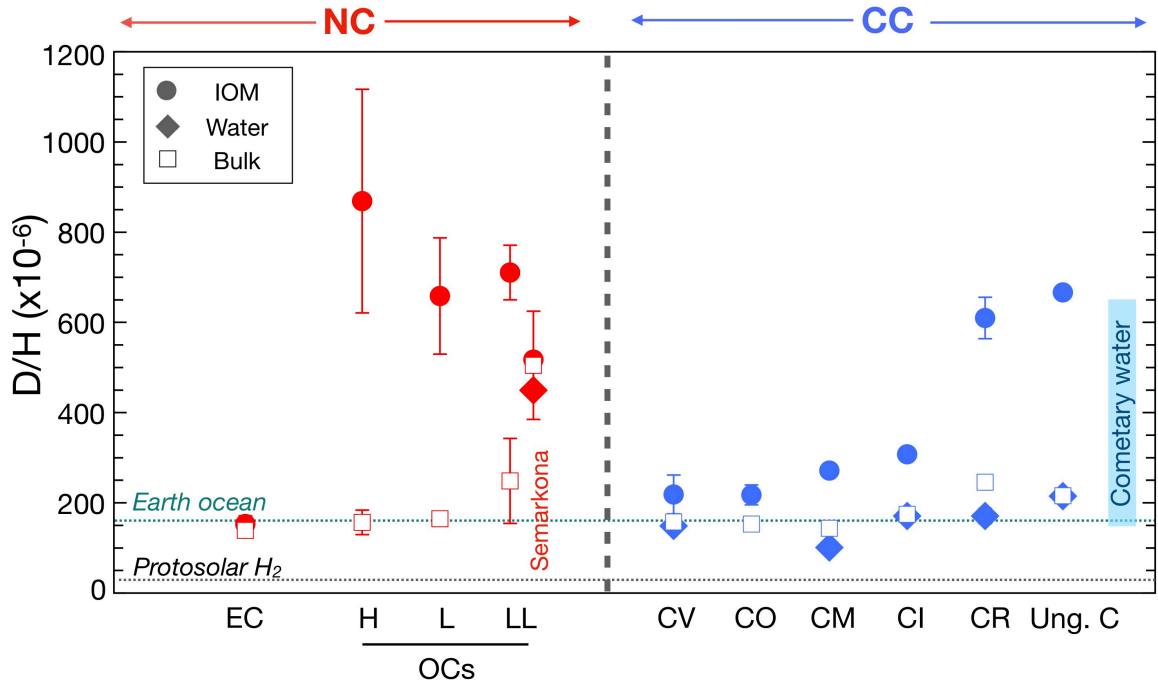

**Fig. 4**

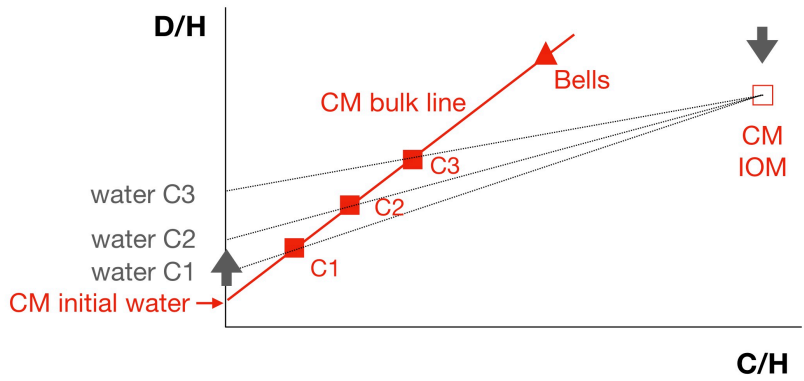

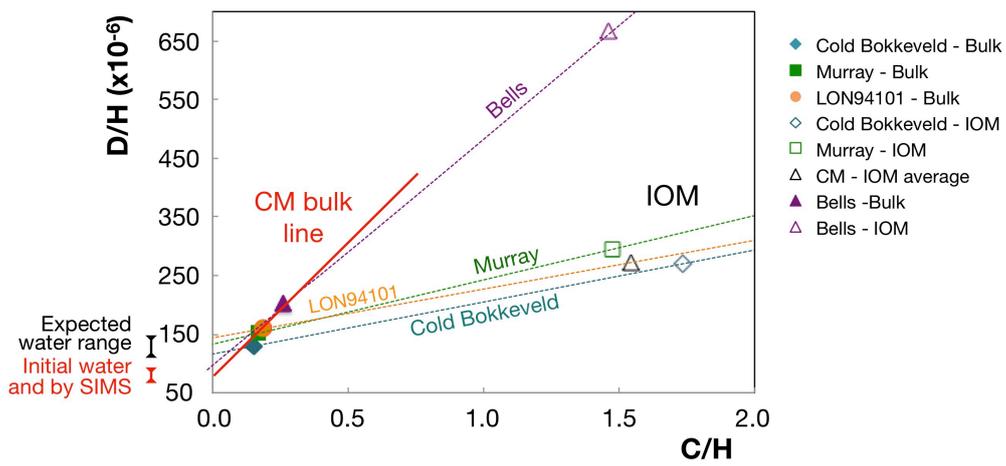

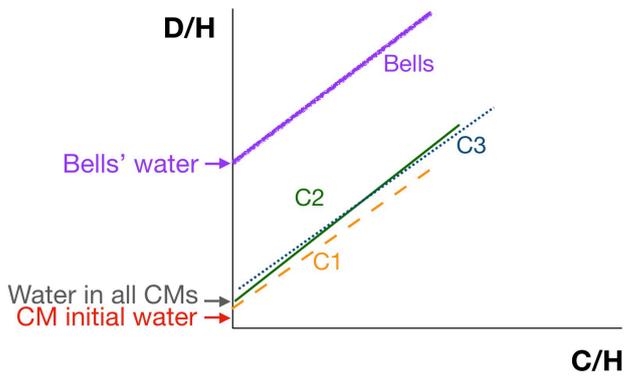

Fig. 5

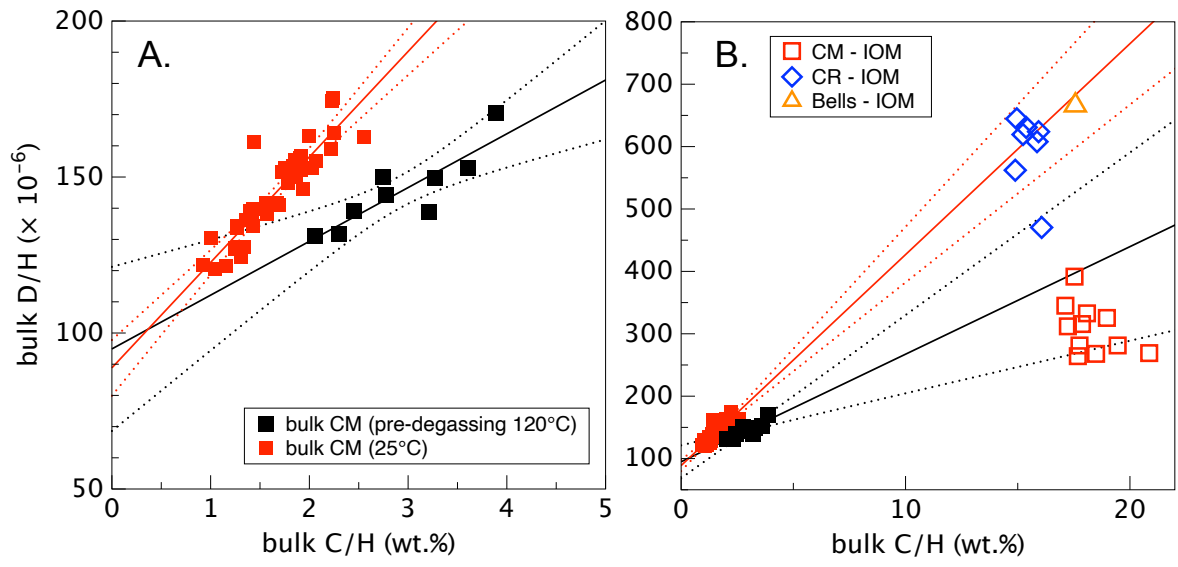

Fig. 6

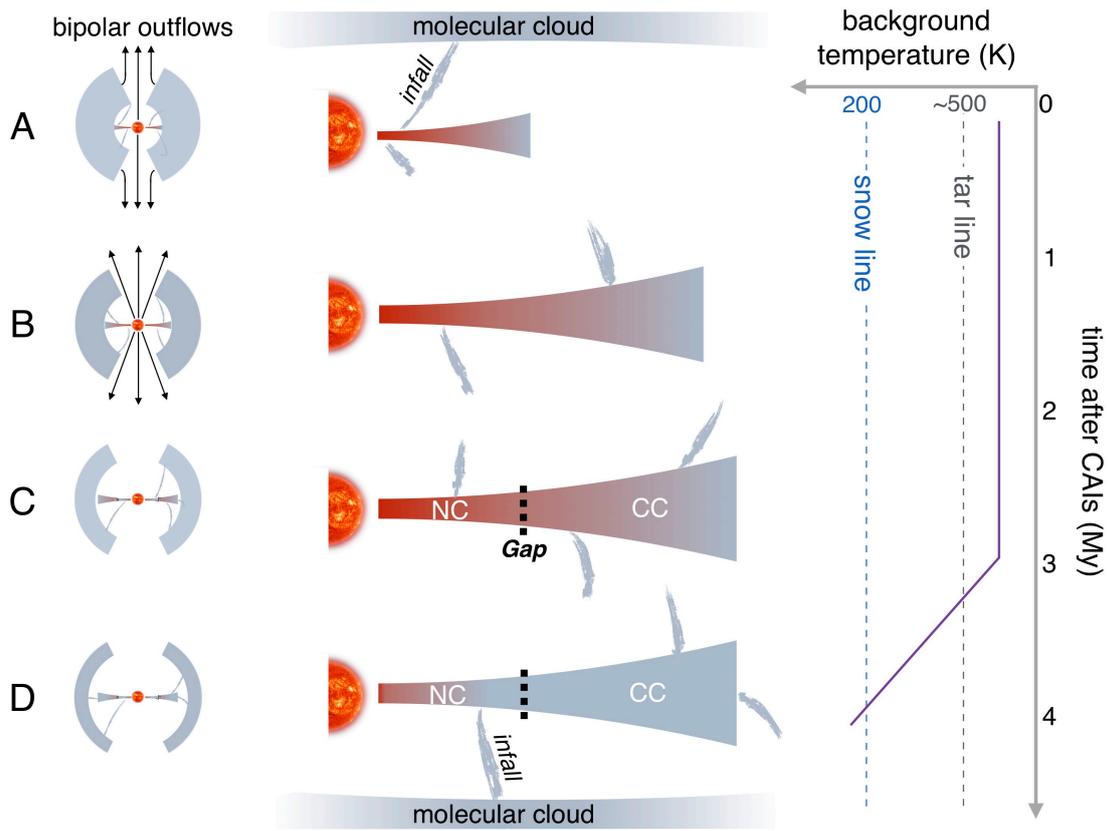

**Fig. 7**

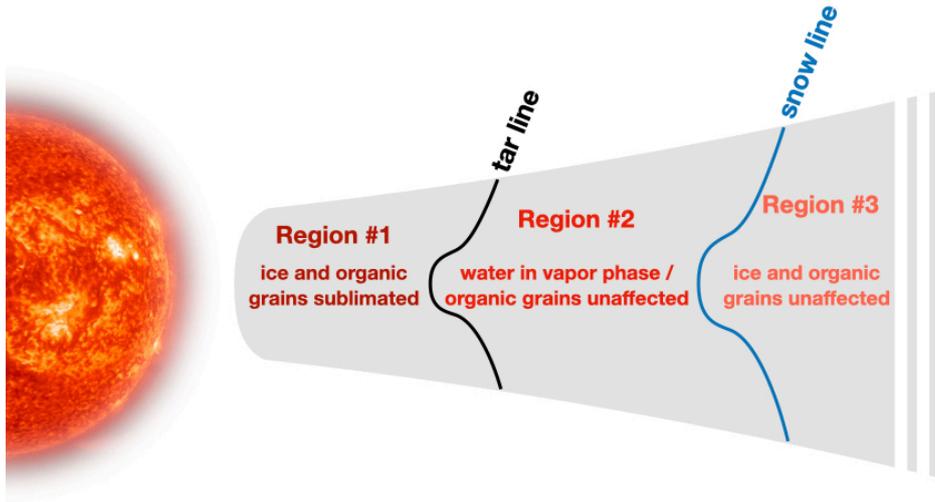

**Fig. 8**

| Meteorite | Type | Session | N | Pearson coeff. | $(D^-/H^-)_0$ ×10$^{-6}$ | error ×10$^{-6}$ | $a$ ×10$^{-4}$ | error ×10$^{-4}$ | $b$ | error | D/H$_{water}$ ×10$^{-6}$ | error ×10$^{-6}$ |
|---|---|---|---|---|---|---|---|---|---|---|---|---|
| Renazzo | CR2 | Dec. 15 | 12 | 0.51 | 202 | 17 | -1.03 | 0.2 | 2.29 | 0.09 | 360 | 32 |
| Maribo | CM2 | Oct. 17 | 25 | 0.77 | 100 | 14 | -1.13 | 0.4 | 2.02 | 0.22 | 90 | 27 |
| LON 94101 | CM2 | Oct. 17 | 22 | 0.78 | 102 | 4 | -1.13 | 0.4 | 2.02 | 0.22 | 94 | 24 |
| Mukundpura | CM2 | Apr. 18 | 24 | 0.68 | 90 | 6 | -0.67 | 0.3 | 1.91 | 0.15 | 105 | 18 |
| Jbilet Wins. | CM2 | Apr. 18 | 35 | 0.33 | 120 | 4 | -0.67 | 0.3 | 1.91 | 0.15 | 162 | 9 |
| Aguas Zarcas | CM2 | July 19 | 21 | 0.55 | 97 | 6 | -1.05 | 0.7 | 2.18 | 0.43 | 107 | 42 |
| DOM08006 | CO3 | Apr. 18 | 14 | 0.33 | 141 | 14 | -0.67 | 0.3 | 1.91 | 0.15 | 203 | 21 |
| Essebi | Ung. C | Apr. 18 | 18 | 0.52 | 150 | 31 | -0.67 | 0.3 | 1.91 | 0.15 | 220 | 46 |
| Bells C | Ung. C | Apr. 18 | 44 | 0.68 | 145 | 12 | -0.67 | 0.3 | 1.91 | 0.15 | 210 | 19 |
| Bells W | Ung. C | Apr. 18 | 32 | 0.73 | 122 | 15 | -0.67 | 0.3 | 1.91 | 0.15 | 167 | 22 |
| Tagish Lake | Ung. C | Apr. 16 | 21 | -0.62 | | | | | | | | |
| Alais | CI | July 19 | 22 | 0.66 | 126 | 11 | -1.05 | 0.7 | 2.18 | 0.43 | 170 | 39 |
| Orgueil | CI | July 19 | 23 | 0.40 | 127 | 16 | -1.05 | 0.7 | 2.18 | 0.43 | 172 | 42 |
| Y-980115 | CY | July 19 | 19 | -0.06 | | | | | | | | |

**Table 1.** Water D/H ratios estimated from SIMS measurements. *N*, number of SIMS analytical spots; Pearson coeff., Pearson correlation coefficient; $(D^-/H^-)_0$, zero intercept of the D/H vs. C/H correlation measured in the chondrite matrix; *a* and *b*, zero intercept and slope of the standard calibration used to estimate the water D/H ratios; and Jbilet Wins., Jbilet Winselwan.